\documentclass{sigchi}

\CopyrightYear{2020}
\setcopyright{acmlicensed}
\doi{https://doi.org/10.1145/3313831.3376145}
\isbn{978-1-4503-6708-0/20/04}
\conferenceinfo{CHI'20,}{April  25--30, 2020, Honolulu, HI, USA}
\acmPrice{\$15.00}

\toappear{\scriptsize Permission to make digital or hard copies of all or part of this work for personal or classroom use is granted without fee provided that copies are not made or distributed for profit or commercial advantage and that copies bear this notice and the full citation on the first page. Copyrights for components of this work owned by others than ACM must be honored. Abstracting with credit is permitted. To copy otherwise, or republish, to post on servers or to redistribute to lists, requires prior specific permission and/or a fee. Request permissions from permissions@acm.org. \\
{\emph{CHI '20, April 25--30, 2020, Honolulu, HI, USA.} } \\
\textcopyright~2020 Association of Computing Machinery. \\
ACM ISBN 978-1-4503-6708-0/20/04\ ...\$15.00. \\
http://dx.doi.org/10.1145/3313831.3376145}

\usepackage{balance}       
\usepackage{graphics}      
\usepackage[T1]{fontenc}   
\usepackage{txfonts}
\usepackage{mathptmx}
\usepackage[pdflang={en-US},pdftex]{hyperref}
\usepackage{color}
\usepackage{booktabs}
\usepackage{textcomp}

\usepackage{microtype}        
\usepackage{bbding}
\usepackage{multirow}
\usepackage{setspace}
\usepackage{url}
\usepackage{todonotes}

\def\plaintitle{``Hey Model!'' -- Natural User Interactions and Agency in Accessible Interactive 3D Models}

\def\emptyauthor{}
\def\plainkeywords{3D printing; Accessibility; Multi-Modal Interaction; Agency}

\makeatletter
\def\url@leostyle{%
  \@ifundefined{selectfont}{
    \def\UrlFont{\sf}
  }{
    \def\UrlFont{\small\bf\ttfamily}
  }}
\makeatother
\def\pprw{8.5in}
\def\pprh{11in}

\definecolor{linkColor}{RGB}{6,125,233}
\hypersetup{%
  pdftitle={\plaintitle},
  pdfauthor={\emptyauthor},
  pdfkeywords={\plainkeywords},
  pdfdisplaydoctitle=true, 
  bookmarksnumbered,
  pdfstartview={FitH},
  colorlinks,
  citecolor=black,
  filecolor=black,
  linkcolor=black,
  urlcolor=linkColor,
  breaklinks=true,
  hypertexnames=false
}

\begin{document}

\title{\plaintitle}

\numberofauthors{3}
\author{%
  \alignauthor{Samuel Reinders\\
    \affaddr{Monash University}\\
    \affaddr{Melbourne, Australia}\\
    \email{Samuel.Reinders@monash.edu}}\\
  \alignauthor{Matthew Butler\\
    \affaddr{Monash University}\\
    \affaddr{Melbourne, Australia}\\
    \email{Matthew.Butler@monash.edu}}\\
  \alignauthor{Kim Marriott\\
    \affaddr{Monash University}\\
    \affaddr{Melbourne, Australia}\\
    \email{Kim.Marriott@monash.edu}}\\
}

\maketitle

\begin{abstract}
While developments in 3D printing have opened up opportunities for improved access to graphical information for people who are blind or have low vision (BLV), they can provide only limited detailed and contextual information. Interactive 3D printed models (I3Ms) that provide audio labels and/or a conversational agent interface potentially overcome this limitation. We conducted a Wizard-of-Oz exploratory study to uncover the multi-modal interaction techniques that BLV people would like to use when exploring I3Ms, and investigated their attitudes towards different levels of model agency. These findings informed the creation of an I3M prototype of the solar system. A second user study with this model revealed a hierarchy of interaction, with BLV users preferring tactile exploration, followed by touch gestures to trigger audio labels, and then natural language to fill in knowledge gaps and confirm understanding.
\end{abstract}

\ccsdesc[500]{Human-centered computing~Accessibility}

\keywords{\plainkeywords}

\printccsdesc

\section{Introduction}

In the last decade there has been widespread interest in the use of 3D printed models to provide blind and low vision (BLV) people access to educational materials~\cite{wedler2012applied}, maps and floor plans~\cite{gual2012visual, Holloway2018}, and for cultural sites~\cite{Rossetti2018}. While these models can contain braille labels this is problematic because of the difficulty of 3D printing braille on a model, the need to introduce braille keys and legends if the labels are too long, and the fact that the majority of BLV people are not fluent braille readers. For this reason, many researchers have investigated interactive 3D printed models (I3Ms) with audio labels~\cite{Gotzelmann2016,Reichinger2016,Giraud2017,Shi2017a,Holloway2018}. 

However, to date almost all research has focused on technologies for interaction, not on ascertaining the needs and desires of the BLV end-user and their preferred interaction strategies. The only research we are aware of that directly addresses this question is that of Shi et. al.~\cite{Shi2017b}. They conducted a Wizard-of-Oz (WoZ) study with 12 BLV participants to elicit preferred user interaction for a pre-defined set of low-level tasks using three simple 3D models. 

Here we describe two user studies that complement and extend this work. We investigate: (1) a wider range of interaction modalities including the use of embodied  conversational agents; (2) the desired level of model agency -- should the model only respond to explicit user interaction or should it proactively help the user; and (3) interaction with more complex models containing removable parts. Such models are common in STEM, e.g. anatomical models in which the organs are removable. We were interested in the interactions and level of model intervention desired by participants, particularly when reassembling the model.

Our first study, Study 1, was an open-ended WoZ study of I3Ms with eight BLV participants. It extends that of~\cite{Shi2017b} in three significant ways. First, it uses a wider variety of models, some of which contain multiple components.  Second, we asked the participants to use the model in any way they wished. This allowed us to elicit a broader range of input and output modalities and interactions. Third, each participant was presented with a low-and high-agency model allowing us to explore the impact of agency.

Study 2 was a follow-up study with six BLV participants which involved exploring a prototype I3M of the solar system, the design of which was informed by Study 1. It supported tactile exploration, tap controlled audio labels, and a conversational interface supporting natural language questions. This study allowed us to confirm the findings of Study 1, whilst addressing a limitation of that study, where participant behaviour may have been biased due to a human providing audio feedback on behalf of the model. While synthesised audio output was actually controlled by the experimenter in Study 2, participants were unaware of this and believed the model was behaving autonomously.

\noindent
\textbf{Contributions}:
Our findings contribute to the understanding of the design space for I3Ms. In particular, we found that: 
\begin{itemize}
    \item \textbf{Interaction modalities:} Participants wished to use a mix of tactile exploration, touch triggered passive audio labels, and natural language questions to obtain information from the model. They mainly wanted audio output but vibration was also suggested. 
   \item \textbf{Independence and model-agency:} Participants wished to be as independent as possible and establish their own interpretations. They wanted to initiate interactions with the model and generally preferred lower model agency, however they did want the model to intervene if they did something wrong such as placing a component in the wrong place.
   \item \textbf{Conversational agent:} Participants preferred more intelligent models that support natural language questions which, when appropriate, could provide guidance to the user.
   \item \textbf{Interaction strategy:} We found a hierarchy of interaction modality. Most participants preferred to glean information and answer questions using tactile exploration, to then use touch triggered audio labels for specific details, and finally use natural language questions to obtain information not in the label or to confirm their understanding.
   \item \textbf{Prior experience:} Interaction choices were driven by participants' prior tactile and technological experiences.
   \item \textbf{Multi-component models:} Participants found models with multiple parts engaging and would remove parts to more readily compare them. 
\end{itemize}

\section{Related Work}

\noindent \textbf{Accessible Graphics \& 3D Models:}
The prevalent methods to produce tactile accessible graphics include using braille embossers, printing onto micro-capsule swell paper, or using thermoform moulds~\cite{Rowell2003}. Their main limitation is that they cannot appropriately convey height or depth~\cite{Holloway2018}, restricting the types of graphics that can be produced to those that are largely flat and two-dimensional in nature. 

As a consequence, handmade models are sometimes used in STEM education and other disciplines that rely on concepts and information that is more three-dimensional in nature. However, while they are uncommon due to difficulties in production and the costs involved~\cite{Holloway2018}, commodity 3D printing has seen the cost and effort required to produce 3D models fall in line with tactile graphics. 3D printing has been used to create accessible models in many contexts: resources to support special education~\cite{Buehler2016}; tangible aids illustrating graphic design theory for BLV students~\cite{McDonald2014}; graphs to teach mathematical concepts~\cite{Brown2012,Hu2015}; programming course curriculum~\cite{Kane2014}; and 3D printed children's books~\cite{kim2015,Stangl2015}.

While 3D printing allows a broader range of accessible graphics to be produced, the low fidelity of 3D printed braille~\cite{Taylor2015,Shi2016} limits the amount of contextual information that can be conveyed on these models, and the updating of braille labels requires model reprinting. Thus, as for tactile graphics, the use of braille labels is problematic, especially considering that the majority of BLV people cannot read braille~\cite{NFIB2009}.

\noindent \textbf{Interactive 3D Printed Models (I3Ms):}
In order to overcome labelling limitations and to make more engaging accessible graphics, 3D printed models have been paired with low-cost electronics and smart devices to create interactive 3D printed models (I3Ms). The majority of studies, however, have focused on technological feasibility rather than usability. 

G\"{o}tzelmann~\cite{Gotzelmann2014} created a smartphone application that was capable of detecting when a user pointed their finger at areas of 3D printed maps during tactile exploration, triggering auditory labels. This method only allowed the use of one hand to tactually explore the 3D prints as it required the user to hold and point their smartphone camera at the print. Shi et. al.~\cite{Shi2016} created Tickers, small percussion instruments that when added to 3D prints can be strummed and detected by a smartphone that triggers auditory descriptions. Testing, however, found that as strummers distorted the model appearance, it interfered with tactile exploration. Further work by Shi et. al.~\cite{Shi2017a} investigated how computer vision can be used to allow BLV people to freely explore and extract auditory labels from 3D prints, but this required affixed 3D markers to support tracking.

Very little research has investigated how BLV users would like to interact with I3Ms, with most studies offering only basic touch interaction~\cite{Taylor2015,Gotzelmann2016,Holloway2018}. A notable exception is Shi et. al.~\cite{Shi2017b} who conducted a WoZ study to examine input preferences and identified distinct techniques across three modalities: gestures; speech; and buttons. These findings were of considerable value; however, the study considered only three simple models, none of which featured detachable components, and focused on a pre-defined set of six generic low-level tasks involving information retrieval and audio note recording. Our research extends this by considering more complex, multi-component models, conversational agents and the impact of model agency.

The role that auditory output can serve in I3Ms is also under-explored, with the majority of I3M research considering only passive auditory labels such as descriptions~\cite{Gotzelmann2016,Reichinger2016,Giraud2017,Shi2017a,Holloway2018} and soundscape cues to add contextual detail~\cite{Hamid2013,Brule2016,Shi2019}. Holloway et. al.~\cite{Holloway2018} gave preliminary guidelines to inform how auditory labels should be used in I3Ms, identifying that: a) trigger points should not distort model appearance, b) triggering should be a deliberate action, and that (c) different gestures should be used to provide different levels of information. Co-designing I3Ms with teachers of BLV students, participants in Shi et. al.~\cite{Shi2019} suggested that in addition to providing passive auditory labelling, I3Ms should allow users to ask the model questions about what it represents. However, this was not explored further in the Shi et. al.~\cite{Shi2019} study. Doing so is a major contribution of our work.

\noindent \textbf{Conversational and Intelligent Agents:}
Allowing the user to ask questions in natural language is a different kind of interface to that usually considered for I3Ms. Advances in voice recognition and natural language processing have resulted in the widespread use of intelligent agents. In particular, intelligent conversational interfaces are increasingly used in everyday life. They include: Siri (Apple), Alexa (Amazon) and Google Assistant. Conversational interfaces have been studied in a variety of contexts, ranging from: health care~\cite{Laranjo2018}, aging~\cite{Nikitina2018}, physical activity~\cite{Watson2012,Bickmore2013}, and for people with a intellectual disability~\cite{Balasuriya2018}. 

With research finding that BLV people find voice interaction convenient~\cite{Azenkot2013}, it comes as no surprise that adoption rates of devices that contain conversational agents, most of which support voice control and text input, is high amongst BLV people. Exploring the use of such devices with 16 BLV participants, Pradhan et al.~\cite{Pradhan2018} identified that 25\% owned at least one device that included a conversational interface, and more broadly found that 15\% of submitted reviews for Amazon Alexa-based devices described use by a person with a disability.

The functions offered through conversational agents are largely passive in nature, requiring the user to activate the agent and to request that a task be performed. However, advances in `intelligent' agents mean that a device may be able to take on a more proactive role. Some agents support pedagogic functions to assist the learning experience by emulating human-human social interaction ~\cite{Moreno2001,Schroeder2013}. Social agency theory suggests that when learners are presented with a human-human social interaction that they become more engaged in the learning environment~\cite{Moreno2001,Mayer2003}. An I3M with these characteristics could facilitate deeper engagement with the model and its wider context. To our knowledge, the integration of agents with higher levels of agency, including the capacity to act and intervene, into accessible 3D printed models has not been previously considered. It is also important to understand how increased agency in an I3M would be accepted, and if it may even lead to a feeling of decreased agency by the end user.

\noindent \textbf{Multi-Modal Interfaces:}
The integration of multiple modalities, such as those mentioned above, can result in interfaces that are capable of communicating richer resolutions of information. When designing interfaces for BLV people this is necessary as other senses must be able to compensate for absence of vision. But Edwards et al.~\cite{Edwards2015} described a `bandwidth problem', wherein other senses are unable to match the capacity of vision, that is unless multiple non-visual senses are utilised. In this context, multi-modal approaches have been used in the creation of assistive aids and tools that use combinations of tactile models with auditory output~\cite{Poppinga2011,Hamid2013}, haptic feedback~\cite{Petrie2002,Goncu2011}, visual feedback for those with residual vision~\cite{Gotzelmann2016}, and olfactory and gustatory perception~\cite{Brule2016}.

Utilising multiple modalities also increases the adaptability of an interface~\cite{Reeves2004}, allowing a user to choose what modalities they want to interact with based upon context or ability. A user may be uncomfortable utilising an interface capable of speech input and output in a noisy environment due to detection problems, or because of privacy concerns~\cite{Abdolrahmani2018}, and may instead prefer another modality such as text input/output, while another user may not have the motor skills to perform gestural input and would instead choose speech input. Our study aims to create I3Ms that allow BLV people to choose their preferred modality where possible, and to uncover any variables that might impact the choice of modalities offered.

\noindent \textbf{Wizard-of-Oz Experiment:} 
When creating new user interfaces, testing can take place before an interface is fully developed. One such method, Wizard-of-Oz (WoZ), used in our two studies, involves an end user interacting with an interface which to some degree is being operated by a `wizard' providing functionality that is yet to be fully implemented~\cite{kelley1984}. By design WoZ typically involves some level of deception, often by omission, where end users may not be aware that they are interacting with an incomplete interface. WoZ methods have been used extensively within HCI research including  the development of conversational agents~\cite{vtyurina2018}, display interfaces~\cite{akers2006} and human-robot interaction~\cite{kahn2008}. 

Within the context of interfaces designed for BLV people, in addition to the work of Shi et. al.~\cite{Shi2017b}, WoZ has been used to explore non-visual web access~\cite{ashok2014}, smartwatch interfaces~\cite{billah2018} and social assistive aids~\cite{rader2014}. While this participant group may be seen as more vulnerable to the illusion that a WoZ interface is fully implemented, many studies with BLV participants explicitly state that the involvement of some deception is integral to their WoZ experiment~\cite{ashok2014,billah2018,rader2014}. Our studies stay within acknowledged WoZ methods, with the true nature of the WoZ deception revealed to participants either at the start (Study 1) or conclusion (Study 2) to ensure transparency.

\section{Study 1: Initial Exploration}
Our first study aimed to better understand which interaction strategies are most natural for BLV users of I3Ms and their preferred level of model agency. We employed a WoZ methodology with a `wizard' providing auditory output on behalf of the model.  WoZ allowed participants to interact with 3D models in any way they felt natural,  while free of technological constraints and for the researchers to readily manipulate the model's degree of agency.

\subsection{Participants}
Study 1 was undertaken with eight participants. They were recruited through the disability support services office of the researchers' home university campus, and through mailing lists of BLV support and advocate groups. Demographic information is summarised in Table~\ref{tab:table1}.

\begin{table}[!htbp]
\centering
\caption{Participant demographic information}
\label{tab:table1}
\begin{tabular}{|l|l|l|l|l|l|l|l|l|}
\hline
\textbf{Participant} & \textbf{P1} & \textbf{P2} & \textbf{P3} & \textbf{P4} & \textbf{P5} & \textbf{P6} & \textbf{P7} & \textbf{P8} \\ \hline
\multicolumn{9}{|l|}{\textbf{Level of Vision:}} \\ \hline
Legally Blind & \checkmark & & \checkmark & & & & & \\ \hline
Totally Blind & & \checkmark & & \checkmark & \checkmark & \checkmark &
\checkmark & \checkmark \\ \hline
\multicolumn{9}{|l|}{\textbf{Formats Used:}} \\ \hline
Braille & \checkmark & \checkmark & \checkmark & \checkmark & \checkmark & & \checkmark & \checkmark \\ \hline
Audio & \checkmark & \checkmark & \checkmark & \checkmark & \checkmark & \checkmark & \checkmark & \checkmark \\ \hline
Raised Line & \checkmark & \checkmark & \checkmark & \checkmark & \checkmark & & & \checkmark \\ \hline
3D Models & & \checkmark & \checkmark & \checkmark & \checkmark & & & \\ \hline
\multicolumn{9}{|l|}{\textbf{Familiarity (/4):}} \\ \hline
Tactile Graphics & 4 & 4 & 3 & 4 & 4 & 1 & 3 & 2 \\ \hline
\multicolumn{9}{|l|}{\textbf{Participation:}} \\ \hline
Study 1 & \checkmark & \checkmark & \checkmark & \checkmark & \checkmark & \checkmark & \checkmark & \checkmark \\ \hline
Study 2 & \checkmark & \checkmark & \checkmark & \checkmark & \checkmark & \checkmark & & \\ \hline 
\end{tabular}
\end{table}

Participant age was evenly spread, from early 20's to early 70's. All participants had smartphones and described using inbuilt conversational interfaces, such as Siri and Google Assistant, to fulfil a variety of tasks including reading and responding to messages, making phone calls, looking up public transit routes and even asking to be told jokes. The experiments took place at a location of convenience for the participant. Each experiment lasted approximately 1.5 to 2 hours.

\subsection{Materials}

\begin{figure*}
\centering
  \includegraphics[width=1.8\columnwidth]{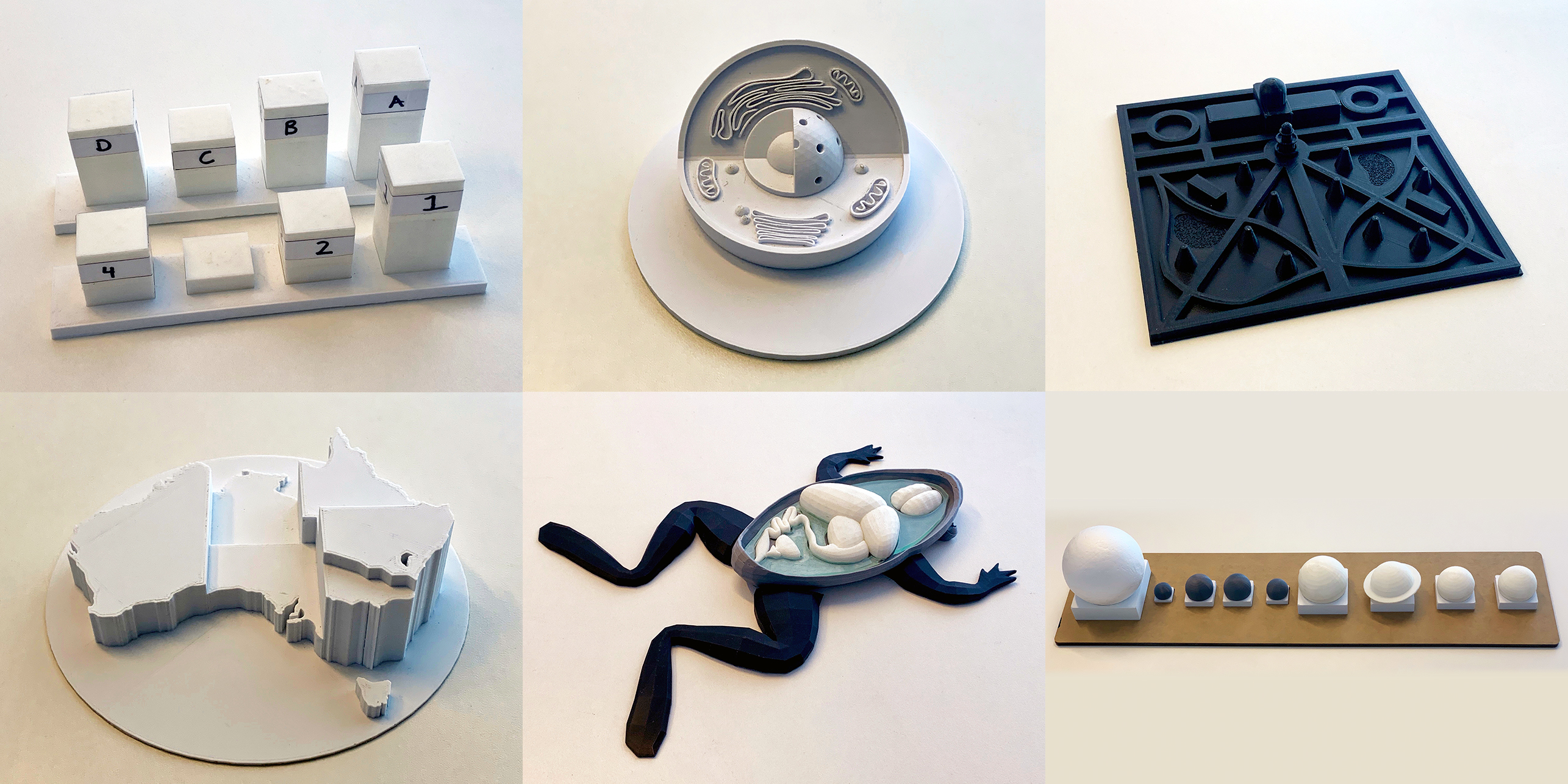}
  \caption{3D models used in Study 1 (left-to-right top-to-bottom): a) Two bar charts of average city temperatures with removable bars; b) Animal cell; c) Park map; d) Thematic map of the population of Australia; e) Dissected frog with removable organs; f) Solar system with removable planets}~\label{fig:figure1}
\end{figure*}

Six 3D printed models representative of materials employed in the classroom or used in Orientation and Mobility (O\&M) training were selected for use in the study. Potential models were chosen from a shortlist by the DIAGRAM 3D working group or designed with advice sought from the Australia \& New Zealand Accessible Graphics Group (ANZAGG). The list of potential models was also informed by the researchers' previous work with the BLV community and BLV educators, with consideration being made of which models are often requested by these stakeholders. 

The six models used were chosen to vary in their application domain, the kinds of tasks they might support, and complexity. Three contained removable components. These were intended to elicit desired interaction behaviours when components were removed or reassembled, as well as to determine whether participants removed components in order to compare them. The models are shown in Figure 1. They were: two bar charts with removable bars, representing the average temperature for each season in two cities; a model of an animal cell\footnote{\url{https://www.thingiverse.com/thing:689381}}; a map of a popular Melbourne public park; a thematic map of Australia, with the height of each state corresponding to population; a model of a dissected frog\footnote{\url{https://www.youmagine.com/designs/the-frog}}, containing removable organs; and a model of the solar system, including removable planets.

\subsection{Procedure}
Each model was explored by at least two different participants, and most models were shown in both a low-agency and a high-agency mode.

Two researchers were present during the experiment: the facilitator and the wizard. At the beginning of the experiment the facilitator explained to participants that they will be given two magical models, one at a time, capable of hearing, seeing, talking, vibrating and sensing touch, and that the models would react accordingly to any way in which they choose to interact with them. It was explained to participants that one of the researchers would take on the role of the wizard. Participants were asked to verbalise any deliberate actions they made using the `think-aloud' protocol~\cite{Wobbrock2009}, allowing the wizard to act on behalf of the model and provide auditory output through speech, and to verbalise other modalities, when relevant. 

When given a model, participants were asked to first explore and identify it in order to familiarise themselves with its features, and then to use the model for any purpose they wished. They were then asked to interact with the model in order to accomplish a predefined task. Tasks were model-specific and included navigating between a southern bench and a building found in the public park map, determining which state or territory has the largest area on the thematic map, and identifying the planet with the lowest density in the solar system model.

After completing the model-specific tasks, participants were asked to design interaction techniques for a number of generic low-level tasks similar to those of~\cite{Shi2017b}: accessing a description of the model; extracting information about a specific component; comparing two components; recording an audio note related to part of the model; and asking the model questions. Throughout these exercises the facilitator observed how the model was being handled, and the wizard acted appropriately to fulfil the interactions. If participants devised any task of their own they were also encouraged to design associated interaction techniques.

Each participant completed the experiment with two models: a low-agency model followed by a high-agency model. When interacting with the low-agency model, feedback was provided by the wizard only when deliberately initiated by the participant, e.g. a participant indicates that when they tap the model they expect an auditory description or haptic vibration. With high-agency models, the wizard played a more proactive role, providing richer assistance including introducing the model when first touched, and intervening if the participant experienced difficulty during the experiment. 

Once both models had been presented, the participant was taken through a semi-structured interview, allowing the facilitator to ask questions about specific interactions the participant used, whether they felt comfortable interacting with the model, and if differing agency capabilities were useful. All interactions were captured on video and dialogue audio recorded. All footage was transcribed, as were tactual interactions. The findings  below are derived from both observation of the participants' interaction with the models, as well as responses to the questions probing their behaviour.

\subsection{Analysis and Results}
An initial set of themes were derived from the experiment protocol. Two researchers independently conducted the initial coding of a data subset to arrive at the main set of themes, and then independently coded all data based on agreed themes. After the data was coded by the researchers, they met to confirm the coding and reconcile any data that was coded differently. Formal quantitative measures were not undertaken as the coding focus was to extract themes in order to inform design principles. Themes were then consolidated into a final set as presented below.

\subsubsection{Preferred Interaction Techniques}
All participants explored the models extensively with touch, at times for long stretches without any questioning or gestures. This was especially common in the initial exploration, and was the dominant technique for identifying what the model was. As participants became either more comfortable or more confident with the model, other interaction techniques emerged.

Specific gestures were observed being used by six participants, often without the participant realising they were doing them. In particular the gesture of double tapping on an element of the model was carried out to either explicitly expect a response, or in conjunction with a question to the model. One participant chose not to perform gestures, indicating that while they were capable of performing them, that others may not based on physical ability.

Voice commands were used by all participants. After initial tactile exploration was conducted, these were typically used to fill in gaps or clarify their understanding of specific features of the model. Dialogue was usually directed at the model in the form of a specific question.

Two participants, however, began to interact with the model in a more complex way in which multiple modalities were combined seamlessly, such as combining touch with conversationally-provided detail. This appeared to be task-related, as both participants completed the route-finding task on the park map. P1 expressed this: \emph{``Okay so I am going to start at the southern most edge [participant taps the south entry to the gardens], and being a magical model um... can you give me directions to the southern most bench?'' [P1]}.

\subsubsection{Preferred Output Modalities}
The overwhelming preference of participants was to extract information using only touch. This included understanding what the model was, its sub-components, and for comparing sub-components. Verbal output was expected in response to the vast majority of questions asked of the model. Interactions such as those with a conversational agent were dominant, with the user gesturing or speaking to the model and expecting a verbal response in return.

While touch input and verbal output were most common, three participants also wished for haptic output. This was used as a way to quickly identify or find a specific component in the model, e.g. the location on the map or a particular planet in the solar system. Haptics were also raised as a way for the model to provide confirmation, for example to indicate that they were holding the correct sub-component, or had found the destination on a map. One participant asked: \emph{``Can you vibrate when I get to Earth?'' [P4]}.

\subsubsection{Exploration and Task Completion Strategies}
All participants exhibited a well-defined strategy for their initial exploration of the model. While participants had varying degrees of experience with 3D prints and even tactile diagrams, the strategy of scanning model boundaries, and then systematically exploring the different parts or areas was common amongst all participants. When questioned on this, some articulated their strategy clearly while others indicated that they did not have one, even when their behaviours implied that they did.

Strategies to complete tasks varied between participants and tasks, however, it was clear that participants preferred to use only their sense of touch where possible. For example, some tasks required discovery of the largest model sub-part (e.g. planet or bar within the chart), and participants typically would find tactual ways to compare, such as positioning their fingers to feel two bars at once. For this type of task, gestures or verbal questions were typically only used for confirmation.

The presence of removable components in some models also prompted the removal of a part and asking questions of it specifically. In comparison tasks, three participants removed parts and held one in each hand to compare.

\subsubsection{Influence of Prior Experience}
Six participants had significant experience with tactile diagrams, with many also having some experience with 3D models, albeit not as a primary presentation of information for them. This experience appeared to influence both their desire to focus on tactual information gathering early on, as well as their strategies for initial exploration. 

Those who were confident smartphone users tended to use more directed gestures. These included typical gestures such as single, double and triple tap. No gestures were observed that deviated from established smartphone use. Two participants even used the gestures that smartphones use to control voice over functions (e.g. two finger tap), expecting similar behaviours to occur. P2 explained: \emph{``I think that is a good way to go because that knowledge that is out there in iPhone... mainstreaming that kind of technology is a great service to society'' [P2]}.

Participants' experience with conversational agents did not appear to influence verbal interactions, in that it did not make them more or less likely to engage in dialogue with the model. When asked about comfort levels speaking with a model, there did not appear to be a link between experience and comfort. This appeared to be influenced more by where the model exploration was taking place (such as noisy or public environments). A number of the braille readers indicated that they would still like to have either braille on the model, a braille key, or braille instructions to use.

\subsubsection{Design Choices}
The choice of model content was a key aspect in stimulating questioning of the model. Questions would relate to things present (e.g. defined aspects that needed explaining) or elements that were absent (e.g. geographical landmarks on the thematic map). These findings support the guidelines presented by~\cite{Shi2017b} regarding ``Improving Tactile Information'' and the importance of considering the tactual elements of the model itself, especially to promote inquiry. 

The use of removable parts where appropriate proved to be a key design choice. The parts supported comparison (such as the size of planets), and also made for more compelling experiences, \emph{``Being able to pick them up? Yeah I liked it ... I like to be able to hold them in terms of the density and size, it is a bit hard to tell when you don't pick them up''} [P2]. 

Tactile sensations mattered to many of the participants. Model elements such as slight imperfections in the 3D printing process were often identified and either the model or researcher was questioned about them. Conversely, explicit differences in materials used (such as softer filament for planets or frog internals) were not commented on by seven participants. One participant, P8, was able to readily detect material differences between planets and this prompted contextual questions exploring why this was the case. Their use was somewhat subtle in the presented models, so differences may need to be more extreme to promote inquiry. 

Some design choices caused confusions. Most notable was the lack of an equator on the Earth in the solar system model, and the inclusion of Saturn's ring, which was confused for the Earth's equator.
 
\subsubsection{Agency and Independence}
In general, independence was highly valued by participants. Independence emerged in two forms: independence of \textbf{control}; and independence of \textbf{interpretation}. 

Regarding control, typically a participant wanted only to know how to interact with the model and then to explore independently, without model intervention. This was especially true for initial model exploration, \emph{``Look if it were a choice between it being over-helpful and having to ask for help, I would prefer to have to ask it for help than to have it just shouting at me that I am wrong'' [P1]}. Independence of interpretation was evident by the reluctance of the participants to simply ask the model for information or the answer to a task, and a preference to use  touch to allow them to build their own understanding, \emph{``I do like to sort of work through it myself and then sort of reach out if I need the help'' [P3]}. 

Opinion was divided on proactive model identification. Half of the participants seemed to like the chance to explore for themselves before being told by or asking the model what it represented, although when prompted some participants indicated a simple introduction by the model would be of benefit, \emph{``I would like to look at it first, and get an idea of what it is, which is kind of what I did, before I found out what it is meant to be.'' [P4]}. Participants indicated that if they did something `wrong', such as place a part in the wrong spot, or have the model oriented in an inappropriate direction, then proactive model intervention would be appreciated.  

Participants appeared to prefer formal dialogue for delivery of basic facts, and conversational for task solving and more exploratory activity. Some participants felt this increased the notion of engagement with the model, \emph{``I don't know,`I am Earth',  `I am this, I am that', I quite like that, because it means you're interacting with it more...'' [P2]}. Some participants also embodied a character or personality into the model. The expression `Mr Model' was used consistently by one participant when asking questions, and others used variations of that. Similarly, some embodied physical features onto some elements of the model. For example, with the solar system model two participants commented on their hesitance to handle the sun, with P4 speaking: \emph{``So, `Mr Model, is this very very hot?' Ah I thought it was hot, I thought it felt a bit warm!'' [P4]}.

\subsection{Discussion}
The results presented above have direct implications for the design of I3Ms.

\textbf{Interaction modalities:}
There are three clear modalities and interaction types: tactile with no model output; touch gesture with model output; and verbal with model output.  The model was expected to provide verbal or haptic output.  The greatest reliance was on touch. Information that was of a more defined nature (e.g. what an element is, or a basic trivial fact) was typically obtained through a gesture (such as double tap), whereas information that was more complex or was not about a specific element of the model tended to be obtained through a question directly aimed at the model. At times this question was preceded by a tapping of a model sub-component (typically if it related to a particular part of the model), and other times it was simply a question directed generally at the model. Thus, the model should provide true multi-modal interaction allowing the user to indicate the objects of interest with touch and natural language to query these. 

\textbf{Conversational agent:}
When interacting verbally with the model, participants treated the model as a conversational agent. As suggested above, users shifted to verbal interaction when seeking more detailed or complex information. The expression `Mr Model' that was used by one participant is emblematic of the nature of dialogue that took place with the model. This is very consistent with formal greetings used to initiate dialogue with conversational agents and should influence the design of I3Ms.

\textbf{Independence and model-agency:}
Participants strongly desired both independence of control and independence of interpretation. Participants did not go for the `easy option' of simply asking the model for the answer to every question, rather they sought to discover the information through tactual means and deliberate triggering of information held by the model. Model intervention to correct their understanding or incorrect placement of a sub-component were exceptions. Thus, I3Ms should employ a low-level of model agency except when aiding the user if they are having trouble, typically with the removable parts. 

\textbf{Prior experience:}
Participants' prior experience appears to  influence their choice of interactions. This is not only the influence of previous experiences with technologies such as smartphones and conversational interfaces, but also their tactile information gathering experiences more broadly. As such, any gestures implemented must align with touch interface standards. Verbal interactions that take place with conversational agents are often more open and as such have less influence on the design of I3Ms.

\textbf{Multi-component models:}
A novel aspect of this study was the inclusion of 3D models with removable parts. While some participants had trouble with reassembly for the more complex frog model, all participants valued their inclusion and none exhibited or mentioned any discomfort during reassembly. These proved to be a valuable design choice, as they promoted greater levels of interaction, engagement and also inquiry of the model. 

\section{Study 2: Validation}
A second study focused on evaluating a more fully functional I3M that incorporated and validated the interaction techniques and agent functionality identified in Study 1.

\begin{figure}
\centering
  \includegraphics[width=\columnwidth]{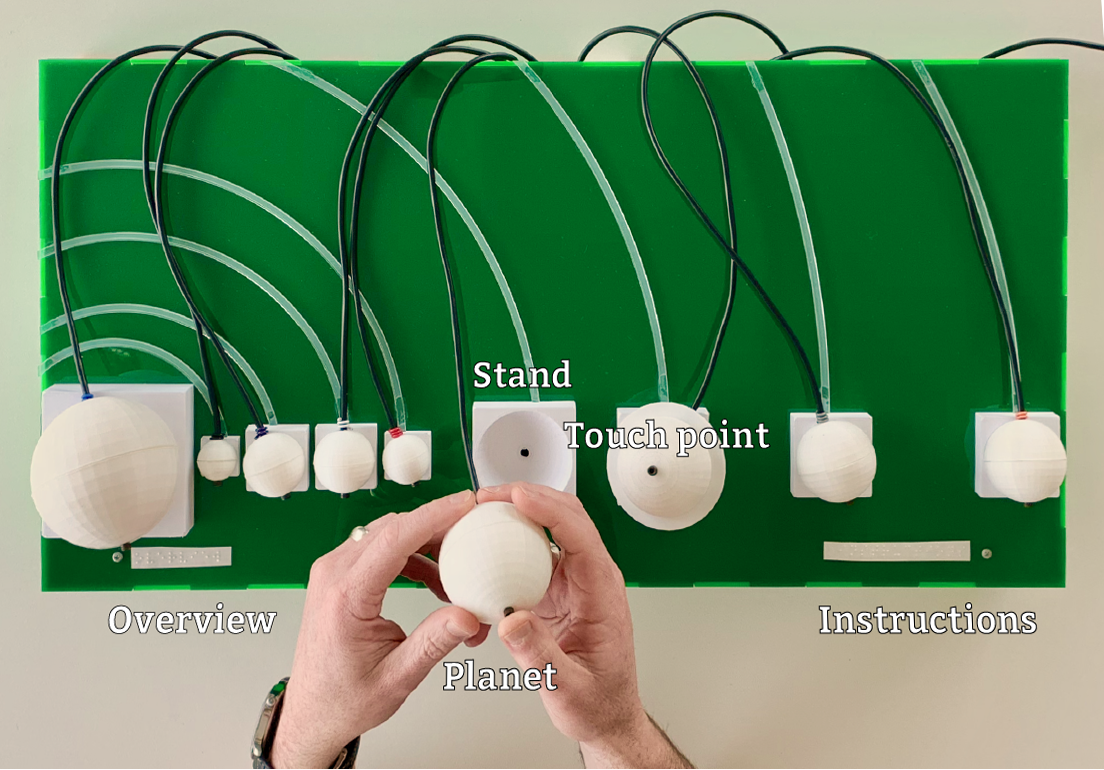}
  \caption{Interactive solar system model used in Study 2}~\label{fig:figure2}
\end{figure}

\subsection{Prototype I3M of the solar system}
The solar system model was selected to undergo further prototyping in the creation of the I3M instantiation. This model was chosen for a number of reasons, including that it allowed for the widest variety of interaction strategies and modalities to be implemented due to its complexity and number of removable components; the model was enjoyed immensely by all participants who were exposed to it in Study 1; and that previous research has expressed a desire to have higher engagement with accessible STEM materials~\cite{butler2017understanding}. One participant made explicit reference to the fact they had not been able to engage with this content previously due to inaccessibility of materials at school. Indeed there seemed to be a knowledge gap regarding the solar system with most of the participants who used this model in Study 1. Based on the results of Study 1, it was determined that the prototype would support the following functionality and behaviours:'

\begin{itemize}
    \item \textbf{Tapping gestures to extract auditory information}: Deliberate gestures such as single tap to trigger component name and double tap to trigger pre-recorded component description audio labels (aligning with standard gesture interactions).
    \item \textbf{Optional overview}: Ability to obtain an overview of the available model interactions, as well as a formal model introduction by tapping dedicated model touch points.
    \item \textbf{On/Off functionality}: Capability to turn auditory output of the model off and on using voice commands to suit the exploratory needs of the user (supporting independence).
    \item \textbf{Braille labelling}: Labels identifying how to access an auditory overview of the prototype and instructions on how to interact with the model (supporting prior knowledge and information gathering techniques).
    \item \textbf{Conversational agent interface}: Ability to ask questions by performing long tap on one or more components or to use an activation phrase (`Hey Model!'), and for the model to respond accordingly to user questions (aligning with standard voice interactions).
    \item \textbf{Model intervention}: The model would assist the user if during interactions an incorrect action was performed (supporting preferred model intervention).
\end{itemize}
The model did not support vibratory feedback due to implementation considerations.

In order to support these functions, a single-board computer was paired with a capacitive touch board containing 12 touch sensors. Each touch sensor was wired to a screw which acted as a touch point. Nine of the touch points were embedded in 3D printed models that represented the Sun and the eight planets of the solar system. The electronics were mounted inside an enclosure constructed out of laser cut acrylic sheets, with stands 3D printed and attached to the top of the enclosure to seat the planets in their correct order. Each planet was tethered using an insulated cable, allowing planets to be removed from their stand while still functioning. A software library was modified to detect when a single, double or long tap was performed on each touch point, and to trigger the playback of an associated audio label through a connected speaker. A single tap would trigger a recording of the component's name, and a double tap a recording of descriptive information, which for the planets included the following: planet name, order from the Sun, radius, type of planet (terrestrial/gas) and composition.

Two additional touch points were placed next to braille labels at the front of the model that read `overview' and `instruction'. The twelfth touch point was embedded in an additional 3D printed cube that was to be used as a training device. 

\subsection{Procedure}
Six of the eight participants from Study 1 were available to take part in Study 2. Using the same participants across both studies was a deliberate methodological choice, as Study 2 focused on validating the findings of Study 1 by having participants interact with an instantiation that supported the preferences and behaviours observed in Study 1. Study 2 was a partial WoZ study, in which the prototype model provided most of the core functionality with the exception of the conversational interface. This was done to validate this functionality before fully implementing it, and was provided using text-to-speech generated on the fly by the researcher. Responses to questions were generated using the same synthesised speech engine as the audio labels embedded into the model. Participants believed the model was behaving autonomously, however for transparency, the true nature of the WoZ implementation was revealed to participants at the conclusion of the study. 

Participants were asked to explore the prototype model and undertake a number of researcher-directed tasks. For consistency between studies, the participant was afforded considerable time to explore the model before a series of guided tasks were presented. In contrast to Study 1, participants were trained in the use of the model, as the purpose of this study was not to identify natural interactions, but rather to validate those previously identified. Furthermore, the user was given more directed tasks to complete with the model. This was to prompt the participant to have opportunity to utilise the full suite of features. The tasks that each user was required to complete were:

\begin{itemize}
    \item \textbf{Simple information gathering}: Exploring and identifying the model (\textbf{T1}), the order of the planets from the Sun (\textbf{T2}), and which of the planets were gas giants (\textbf{T3}).
    \item \textbf{Comparison tasks}: Finding out the radius of two planets (\textbf{T4}), identifying the largest planet (\textbf{T5}), and which planet has the longest orbit (\textbf{T6}).
    \item \textbf{Complex question answering}: Using information from T6 to establish relationship between planet distance and orbital period (\textbf{T7}).
    \item \textbf{Model reassembly}: Placing four planets back into their stands while the model confirms correct or incorrect placement (\textbf{T8}).
\end{itemize}

In order to force the use of different interaction modalities, T3 and T4 could not be answered using only touch, and T6 required the user to ask the model. The study concluded with questions regarding the participants' experiences with the model:

\begin{itemize}
    \item General questions regarding \textbf{engagement with the model}, including: whether they enjoyed using the model, whether they learnt anything, and whether they thought similar models would have been useful during their education.
    \item Preference of \textbf{interaction techniques} and how they aligned with Study 1, including: why they gravitated towards a particular method of interaction, and whether there were any additional interactions they would like the model to support.
    \item Questions regarding the level of \textbf{agency of the model}. Elements were derived from the Godspeed questionnaire~\cite{Bartneck2009}, used to measure users' perception of AI and robot interaction. This included: perceived competence and intelligence of the model, and if participants found the level of intervening assistance provided useful.
    \item Satisfaction with the \textbf{level of independence} afforded, and whether they felt in control during interactions, and how this aligned with their interaction technique preferences.
    \item Inquiry into \textbf{participant comfort} when undertaking dialogue and conversing with the model, and self-identified safety and emotions during these interactions.
    \item \textbf{Overall satisfaction} with the design and preference of the model over more traditional graphical representations of information.
\end{itemize}
Where relevant, participants' preferences in Study 1 were raised for direct comparison.

\subsection{Experiment Conditions}
Each of the participants was taught how to use the model with a simple training interface. The interface taught the range of gesture interactions, as well as the two natural language interfaces (long tap and ``Hey Model!''). The order of introduction to these techniques was counterbalanced in order to remove bias. Similarly, some tasks were counterbalanced: Tasks 4 \& 5: `Tell me the Radius?'; and `Which is the biggest?'. The experiments ranged in time from 1 to 2 hours, primarily directed by the levels of engagement from the participant. 

\subsection{Results}
Video and audio was recorded, and all video transcribed. All transcriptions were coded by two researchers, using the same themes as identified in Study 1 for direct comparison and validation of interaction techniques from that stage. In reporting the results, focus is placed on the alignment of preferred input and output modalities, and model agency.

\subsubsection{Task Completion}
Interaction modalities for the tasks are summarised in Table~\ref{tab:table2}.
\begin{itemize}
    \item T1: All participants began by tactually exploring the model before performing tap gestures on the various components of the model that revealed what the model represented. Participants continued interacting with the model, with three activating natural language and asking for further information. \emph{``... it doesn't mention the rings... Hey Model, is there more information on Saturn?'' [P5]}
    \item T2: Five participants tactually explored each planet before using single tap gestures to identify each of them. One participant, P5, chose not to use tap gestures as they felt confident in their knowledge of the planets and instead named each while touching the corresponding part of the model.
    \item T3: Five participants used tap gestures to determine whether each planet was a gas giant, performing a double tap to trigger a planet's audio label. One participant instead simply asked \emph{``Hey Model, can you identify the gas planets?'' [P6]}
    \item T4: Five participants used tap gestures to determine the radius of Uranus and Neptune, performing a double tap to trigger each planet's audio label. P6 again chose to rely upon natural language questioning and asked the model. 
    \item T5: All participants were able to correctly identify that Jupiter was the largest planet using tactual exploration, comparing the size of each planet. Three participants performed double tap gestures to confirm their answers by listening to a planet's audio label, \emph{``I will just double check Saturn which is next... Yep Jupiter is the largest...'' [P3]}
    \item T6: Five participants engaged in natural language questioning in order to determine which planet had the longest orbit time, with one participant narrating  \emph{``Now it hasn't yet given me that so... let's ask... Hey Model, which planet takes the longest to ... orbit the Sun?'' [P4]}. P1 was confident they knew that Neptune had the longest orbit, but was unsure of the exact duration and engaged natural language through a long tap and asked \emph{``What is the orbit time of Neptune?''}
    \item T7: Using information uncovered in T6, four participants were able to establish the relationship between the distance of a planet from the Sun and its time of orbit without interacting with the model. The others queried the model.
    \item T8: All participants started by tactually scanning the empty planet stands to determine sizes. Three participants then searched for the largest unseated planet. All participants performed tap gestures to identify each planet they picked up before inserting it into a stand. Three participants were able to correctly place all four unseated planets without any mistakes, with the remaining three participants requiring model intervention. P6 found the assistance provided by the model useful when placing a planet in the correct stand, stating \emph{``That is what I wanted to know, fabulous''}.
\end{itemize}

\begin{table}[]
\centering
\caption{Applicable modalities for each task and used by participants}
\label{tab:table2}
\begin{tabular}{|l|l|l|l|l|l|l|l|}
    \cline{1-8}
    \multirow{2}{*}{\textbf{Task:}} & \multirow{2}{*}{\textbf{Applicable Modalities:}} & \multicolumn{6}{c|}{\textbf{Modalities Used:}} \\ \cline{3-8} & & \rotatebox{90}{P1} & \rotatebox{90}{P2} & \rotatebox{90}{P3} & \rotatebox{90}{P4} & \rotatebox{90}{P5} & \rotatebox{90}{P6\phantom{.}} \\ \hline
    \multicolumn{1}{|l|}{\multirow{3}{*}{{\textbf{T1}}}} & Touch & \checkmark & \checkmark & \checkmark & \checkmark & \checkmark & \checkmark \\ \cline{2-8}
    \multicolumn{1}{|l|}{} & Tap & \checkmark & \checkmark & \checkmark & \checkmark & \checkmark & \checkmark \\ \cline{2-8} 
    \multicolumn{1}{|l|}{} & Natural Language & \checkmark & \checkmark & & \checkmark & \checkmark & \\ \hline
    \multicolumn{1}{|l|}{\multirow{3}{*}{{\textbf{T2}}}} & Touch & \checkmark & \checkmark & \checkmark & \checkmark & \checkmark & \checkmark \\ \cline{2-8}
    \multicolumn{1}{|l|}{} & Tap & \checkmark & \checkmark & \checkmark & \checkmark & & \checkmark \\ \cline{2-8}
    \multicolumn{1}{|l|}{} & Natural Language & & & & \checkmark & & \\ \hline
    \multicolumn{1}{|l|}{\multirow{2}{*}{{\textbf{T3}}}} & Tap & \checkmark & \checkmark & \checkmark & \checkmark & \checkmark & \\ \cline{2-8}
    \multicolumn{1}{|l|}{} & Natural Language & & & & & & \checkmark \\ \hline
    \multicolumn{1}{|l|}{\multirow{2}{*}{{\textbf{T4}}}} & Tap & \checkmark & \checkmark & \checkmark & \checkmark & \checkmark & \\ \cline{2-8}
    \multicolumn{1}{|l|}{} & Natural Language & & & & & & \checkmark \\ \hline
    \multicolumn{1}{|l|}{\multirow{3}{*}{{\textbf{T5}}}} & Touch & \checkmark & \checkmark & \checkmark & \checkmark & \checkmark & \checkmark \\ \cline{2-8}
    \multicolumn{1}{|l|}{} & Tap & & \checkmark & \checkmark & \checkmark & & \\ \cline{2-8} 
    \multicolumn{1}{|l|}{} & Natural Language & & & & & & \\ \hline
    \multicolumn{1}{|l|}{\multirow{1}{*}{{\textbf{T6}}}} & Natural Language & \checkmark & \checkmark & \checkmark & \checkmark & \checkmark & \checkmark \\ \hline
    \multicolumn{1}{|l|}{\multirow{1}{*}{{\textbf{T7}}}} & Natural Language & & & & \checkmark & & \checkmark \\ \hline
    \multicolumn{1}{|l|}{\multirow{3}{*}{{\textbf{T8}}}} & Touch & \checkmark & \checkmark & \checkmark & \checkmark & \checkmark & \checkmark \\ \cline{2-8}
    \multicolumn{1}{|l|}{} & Tap & \checkmark & \checkmark & \checkmark & \checkmark & \checkmark & \checkmark \\ \cline{2-8} 
    \multicolumn{1}{|l|}{} & Natural Language & & & \checkmark & & & \\ \hline
\end{tabular}
\end{table}

\subsubsection{Post-Task Questions}
\textbf{Engagement with the model}: All six participants indicated that they enjoyed interacting with the model, citing how it gave them \emph{``A better idea of how it all works and everything'' [P4]} to how it was \emph{``Sort of fun because you could move them around and feel the different sizes'' [P3]}. Five participants felt they learnt new knowledge ranging from the general size of the planets, their radii and which were gas giants. All participants agreed that it would have been useful having access to similar models during their education, with P2 highlighting that \emph{``It would have been very engaging, especially in Year 7''} and other participants suggesting possible uses when teaching anatomy, physics and chemistry.

\textbf{Interaction techniques}: Preference of interaction techniques largely aligned with the interaction strategies used in Study 1. All six participants outlined that touch was important, with P2 suggesting that they gravitated towards touch because \emph{``I know it is built to be touched ... touch is very important to me...''}. When extracting information, four participants spoke of how they would move to use natural language when they were unable to extract the information they desired using tap gestures, \emph{``When the information isn't available through tap I like being able to extend that information by being able to ask questions'' [P4]}. Additionally, despite the text-to-speech interface involving a slight delay of a few seconds, all participants believed that the model was behaving autonomously.

\textbf{Agency of the model}: When placing the planets back into their stands, all six participants found the level of intervening assistance provided useful. Two participants said that the model having this level of agency allowed them to complete the task faster. P3 suggested a model's level of agency should be controllable, \emph{``Maybe if you could turn it on or off, because you might not always want it?''}. When asked if they would like the model to have an embodied role, participants were divided, with three suggesting a teaching role may be useful with school children.

\textbf{Level of independence}: All six participants indicated that they largely felt in control of their interactions during their time with the model. This aligned with the desire for an independent experience found in Study 1. P4 spoke of how the conversational interface supported their desire to explore independently, \emph{``[the model has] given to me what a sighted person would give me if they were helping me to look at something, so it makes the whole experience a lot more independent''}. One participant even connected independence and knowing how to interact with the model, suggesting that  \emph{``... showing me the different gestures to start with is very important'' [P2]}.

\textbf{Participant comfort}: Five participants spoke of how they felt comfortable asking the model questions, with two suggesting this may be due to prior experience with voice activation, \emph{``Just as intuitive as asking any other sort of voice assistant questions'' [P1]}. P4 said that they \emph{``Felt confident asking a question, [but weren't] confident in how to phrase questions'' [P4]}, which was echoed by P5. When asked to place how they felt on a scale of agitated to neutral to calm during interactions, all six participants rated their experience as ``calm".

\textbf{Overall satisfaction}: Participants were asked to choose their preference from a) models that only contain braille, to b) those that also output speech through tap gestures, c) models that further understand speech and answer questions when asked or tapped, or d) a conversational interface that answers questions with no attached physical model. There was strong agreement amongst all six participants towards c), which aligned with the solar system prototype. P1 spoke of how \emph{``With a model like this you are able to have more information than with a dumb model''}, while P2 suggested it was \emph{``Catering to all sorts of abilities''}. All participants suggested a physical model was necessary and that the experience wouldn't be as satisfying without one because \emph{``It is not multi-modal and everyone learns better multi-modal'' [P5]} and that physical models making things \emph{``More playful, more fun'' [P2]}

\subsection{Discussion}
The findings from Study 2 aligned strongly with those in Study~1.

\textbf{Interaction modalities:} Table 2 confirms the findings of Study 1, in that multiple modalities are used throughout exploration and task completion. While there is a tendency towards simple touch interaction, all participants used tap-initiated auditory information, as well as natural language.

\textbf{Conversational agent:} Participants felt that the model exhibited some levels of intelligence, and were comfortable engaging in natural language dialogue, even though it was the least preferred modality of interaction. The expression ``Hey Model!'' was also seen as feeling natural when initiating dialogue with the model. 

\textbf{Independence and model-agency:} Study 2 further confirmed participants' wishes to be in control of their experience and assemble their own understandings.

\textbf{Prior experience:} Prior tactile experience emerged as a strong factor in Study 2. Of the six participants, P6 had the least amount of tactile experience, in part due to the age of onset of vision loss as well as opportunity. Participant P6 exhibited a different interaction strategy, using natural language earlier and more often than the others. P6 also engaged in less complex touch exploration. Regarding experience with technology, the use of tap gestures in the prototype was well understood, with only minor sensitivity issues. This also extended to the natural language interface, with multiple participants detailing that the latency involved when asking questions was similar to that experienced when using other conversational agents.

\textbf{Multi-component models:} As with Study 1, having multiple components in the model promoted greater engagement with it, as well as encouraging deeper levels of inquiry. 

\textbf{Interaction strategy:} A key finding of Study 2 was that a clear hierarchy of interaction emerged when BLV people engaged with I3Ms. 
\begin{enumerate}
    \item Tactile Exploration -- It appears that this is seen by participants as supporting the greatest level of independence in both control and interpretation. It is also likely the most familiar information gathering technique for this cohort (i.e. experience with tactile diagrams, etc) and allows the user to come to their own conclusions. 
    \item Gesture-Driven Enquiry -- This is the extraction of information about the model or a component of the model through tapping and other deliberate gestures. This is used to elicit low-level  `factual' information, and is used when it is not possible to obtain this through touch.
    \item Natural Language Interrogation -- While participants are reasonably comfortable asking questions of the model, conversation with the model is largely used to fill gaps of knowledge, or to confirm understanding. 
\end{enumerate}

\section{Conclusion}
In this paper we have presented two user studies investigating blind and low vision (BLV) peoples' preferred interaction techniques and modalities for interactive 3D printed models (I3Ms). Study 1 utilised a Wizard-of-Oz methodology, and Study 2 confirmed the results by evaluating the use of a prototype I3M of the solar system the design of which was informed by the findings of Study 1. 

We found that participants wished to use a mix of tactile exploration, touch triggered passive audio labels, and natural language questioning to obtain information from the model with a mix of audio and haptic output. They enjoyed engaging with models that had multiple parts, and would remove parts to further explore and compare them.

When talking to the model, participants treated it as a conversational agent and indicated that they preferred more intelligent models that support natural language and which, when appropriate, could provide guidance to the user. Participants wished to be as independent as possible and establish their own interpretations. They wanted to initiate interactions with the model and generally preferred lower model agency. However, they did want the model to intervene if they did something wrong such as placing a component in the wrong place. 
  
The desire for independent exploration led to a hierarchy of interaction modalities: most participants preferred to glean information and answer questions using tactile exploration, then to use touch triggered audio labels for specific details, finally using natural language questions to obtain information not in the label or to confirm their understanding. However, interaction choices were driven by participants' prior tactile and technological experiences. Not only are these findings of significance to the assistive technology community, they also have wider implications to HCI. In particular the combination of I3Ms with conversational agents suggests a radically new kind of embodied conversational agent, one that is physically embodied and that can be perceived tactually by a BLV person, rather than the  traditional embodied conversational agent that is perceived visually and has a human-like experience~\cite{cassell2000more}. 

Such physically embodied conversational agents raise many interesting research questions, including their perceived agency, autonomy and acceptance by the end user. There are also many questions to be answered on how such agents can be implemented. A major focus of our future research will be to design and construct a fully functional prototype, conduct more extensive user evaluations with a variety of models, including maps, and to explore whether model agency preferences differ with age and environment.

\section{Acknowledgments}
This research was supported by an Australian Government Research Training Program (RTP) Scholarship. Additionally, we thank all the research participants for their time and expertise.

%
%
%
%
%
\balance{}

\balance{}

\bibliographystyle{SIGCHI-Reference-Format}
\bibliography{references}


\begin{thebibliography}{00}


\ifx \showCODEN    \undefined \def \showCODEN     #1{\unskip}     \fi
\ifx \showDOI      \undefined \def \showDOI       #1{{\tt DOI:}\penalty0{#1}\ }
  \fi
\ifx \showISBNx    \undefined \def \showISBNx     #1{\unskip}     \fi
\ifx \showISBNxiii \undefined \def \showISBNxiii  #1{\unskip}     \fi
\ifx \showISSN     \undefined \def \showISSN      #1{\unskip}     \fi
\ifx \showLCCN     \undefined \def \showLCCN      #1{\unskip}     \fi
\ifx \shownote     \undefined \def \shownote      #1{#1}          \fi
\ifx \showarticletitle \undefined \def \showarticletitle #1{#1}   \fi
\ifx \showURL      \undefined \def \showURL       #1{#1}          \fi

\bibitem{Hamid2013}
{Nazatul~Naquiah Abd~Hamid} {and} {Alistair~D.N. Edwards}. 2013.
\newblock \showarticletitle{Facilitating Route Learning Using Interactive
  Audio-tactile Maps for Blind and Visually Impaired People}. In {\em CHI '13
  Extended Abstracts on Human Factors in Computing Systems} {\em (CHI EA '13)}.
  ACM, New York, NY, USA, 37--42.
\newblock
\showISBNx{978-1-4503-1952-2}
\showDOI{%
\url{http://dx.doi.org/10.1145/2468356.2468364}}


\bibitem{Abdolrahmani2018}
{Ali Abdolrahmani}, {Ravi Kuber}, {and} {Stacy~M. Branham}. 2018.
\newblock \showarticletitle{"Siri Talks at You": An Empirical Investigation of
  Voice-Activated Personal Assistant (VAPA) Usage by Individuals Who Are
  Blind}. In {\em Proceedings of the 20th International ACM SIGACCESS
  Conference on Computers and Accessibility} {\em (ASSETS '18)}. ACM, New York,
  NY, USA, 249--258.
\newblock
\showISBNx{978-1-4503-5650-3}
\showDOI{%
\url{http://dx.doi.org/10.1145/3234695.3236344}}


\bibitem{akers2006}
{David Akers}. 2006.
\newblock \showarticletitle{Wizard of Oz for Participatory Design: Inventing a
  Gestural Interface for 3D Selection of Neural Pathway Estimates}. In {\em CHI
  '06 Extended Abstracts on Human Factors in Computing Systems} {\em (CHI EA
  '06)}. ACM, New York, NY, USA, 454--459.
\newblock
\showISBNx{1-59593-298-4}
\showDOI{%
\url{http://dx.doi.org/10.1145/1125451.1125552}}


\bibitem{ashok2014}
{Vikas Ashok}, {Yevgen Borodin}, {Svetlana Stoyanchev}, {Yuri Puzis}, {and}
  {I.~V. Ramakrishnan}. 2014.
\newblock \showarticletitle{Wizard-of-Oz Evaluation of Speech-driven Web
  Browsing Interface for People with Vision Impairments}. In {\em Proceedings
  of the 11th Web for All Conference} {\em (W4A '14)}. ACM, New York, NY, USA,
  Article 12, 9 pages.
\newblock
\showISBNx{978-1-4503-2651-3}
\showDOI{%
\url{http://dx.doi.org/10.1145/2596695.2596699}}


\bibitem{Azenkot2013}
{Shiri Azenkot} {and} {Nicole~B. Lee}. 2013.
\newblock \showarticletitle{Exploring the Use of Speech Input by Blind People
  on Mobile Devices}. In {\em Proceedings of the 15th International ACM
  SIGACCESS Conference on Computers and Accessibility} {\em (ASSETS '13)}. ACM,
  New York, NY, USA, Article 11, 8 pages.
\newblock
\showISBNx{978-1-4503-2405-2}
\showDOI{%
\url{http://dx.doi.org/10.1145/2513383.2513440}}


\bibitem{Balasuriya2018}
{Saminda~Sundeepa Balasuriya}, {Laurianne Sitbon}, {Andrew~A. Bayor}, {Maria
  Hoogstrate}, {and} {Margot Brereton}. 2018.
\newblock \showarticletitle{Use of Voice Activated Interfaces by People with
  Intellectual Disability}. In {\em Proceedings of the 30th Australian
  Conference on Computer-Human Interaction} {\em (OzCHI '18)}. ACM, New York,
  NY, USA, 102--112.
\newblock
\showISBNx{978-1-4503-6188-0}
\showDOI{%
\url{http://dx.doi.org/10.1145/3292147.3292161}}


\bibitem{Bartneck2009}
{Christoph Bartneck}, {Dana Kuli{\'{c}}}, {Elizabeth Croft}, {and} {Susana
  Zoghbi}. 2009.
\newblock \showarticletitle{Measurement Instruments for the Anthropomorphism,
  Animacy, Likeability, Perceived Intelligence, and Perceived Safety of
  Robots}.
\newblock {\em International Journal of Social Robotics\/} {1}, 1 (01 Jan
  2009), 71--81.
\newblock
\showISSN{1875-4805}
\showDOI{%
\url{http://dx.doi.org/10.1007/s12369-008-0001-3}}


\bibitem{Bickmore2013}
{Timothy~W. Bickmore}, {Daniel Schulman}, {and} {Candace Sidner}. 2013.
\newblock \showarticletitle{Automated interventions for multiple health
  behaviors using conversational agents}.
\newblock {\em Patient Education and Counseling\/} {92}, 2 (2013), 142 -- 148.
\newblock
\showISSN{0738-3991}
\showDOI{%
\url{http://dx.doi.org/https://doi.org/10.1016/j.pec.2013.05.011}}


\bibitem{billah2018}
{Syed~Masum Billah}, {Vikas Ashok}, {and} {IV Ramakrishnan}. 2018.
\newblock \showarticletitle{Write-it-Yourself with the Aid of Smartwatches: A
  Wizard-of-Oz Experiment with Blind People}. In {\em 23rd International
  Conference on Intelligent User Interfaces} {\em (IUI '18)}. ACM, New York,
  NY, USA, 427--431.
\newblock
\showISBNx{978-1-4503-4945-1}
\showDOI{%
\url{http://dx.doi.org/10.1145/3172944.3173005}}


\bibitem{Brown2012}
{Craig Brown} {and} {Amy Hurst}. 2012.
\newblock \showarticletitle{VizTouch: Automatically Generated Tactile
  Visualizations of Coordinate Spaces}. In {\em Proceedings of the Sixth
  International Conference on Tangible, Embedded and Embodied Interaction} {\em
  (TEI '12)}. ACM, New York, NY, USA, 131--138.
\newblock
\showISBNx{978-1-4503-1174-8}
\showDOI{%
\url{http://dx.doi.org/10.1145/2148131.2148160}}


\bibitem{Brule2016}
{Emeline Brule}, {Gilles Bailly}, {Anke Brock}, {Frederic Valentin},
  {Gr{\'e}goire Denis}, {and} {Christophe Jouffrais}. 2016.
\newblock \showarticletitle{MapSense: Multi-Sensory Interactive Maps for
  Children Living with Visual Impairments}. In {\em Proceedings of the 2016 CHI
  Conference on Human Factors in Computing Systems} {\em (CHI '16)}. ACM, New
  York, NY, USA, 445--457.
\newblock
\showISBNx{978-1-4503-3362-7}
\showDOI{%
\url{http://dx.doi.org/10.1145/2858036.2858375}}


\bibitem{Buehler2016}
{Erin Buehler}, {Niara Comrie}, {Megan Hofmann}, {Samantha McDonald}, {and}
  {Amy Hurst}. 2016.
\newblock \showarticletitle{Investigating the Implications of 3D Printing in
  Special Education}.
\newblock {\em ACM Trans. Access. Comput.\/} {8}, 3, Article 11 (March 2016),
  28 pages.
\newblock
\showISSN{1936-7228}
\showDOI{%
\url{http://dx.doi.org/10.1145/2870640}}


\bibitem{butler2017understanding}
{Matthew Butler}, {Leona Holloway}, {Kim Marriott}, {and} {Cagatay Goncu}.
  2017.
\newblock \showarticletitle{Understanding the graphical challenges faced by
  vision-impaired students in Australian universities}.
\newblock {\em Higher Education Research \& Development\/} {36}, 1 (2017),
  59--72.
\newblock
\showDOI{%
\url{http://dx.doi.org/10.1080/07294360.2016.1177001}}


\bibitem{cassell2000more}
{Justine Cassell}. 2000.
\newblock \showarticletitle{More than just another pretty face: Embodied
  conversational interface agents}.
\newblock {\it Commun. ACM} {43}, 4 (2000), 70--78.
\newblock


\bibitem{Edwards2015}
{Alistair D.~N. Edwards}, {Nazatul Naquiah~Abd Hamid}, {and} {Helen Petrie}.
  2015.
\newblock \showarticletitle{Exploring Map Orientation with Interactive
  Audio-Tactile Maps}. In {\em Human-Computer Interaction -- INTERACT 2015},
  {Julio Abascal}, {Simone Barbosa}, {Mirko Fetter}, {Tom Gross}, {Philippe
  Palanque}, {and} {Marco Winckler} (Eds.). Springer International Publishing,
  Cham, 72--79.
\newblock
\showISBNx{978-3-319-22701-6}


\bibitem{Giraud2017}
{St\'{e}phanie Giraud}, {Anke~M Brock}, {Marc J-M Mac\'{e}}, {and} {Christophe
  Jouffrais}. 2017.
\newblock \showarticletitle{Map learning with a 3D printed interactive
  small-scale model: Improvement of space and text memorization in visually
  impaired students}.
\newblock {\em Frontiers in Psychology\/} {8}, 930 (2017).
\newblock
\showDOI{%
\url{http://dx.doi.org/10.3389/fpsyg.2017.00930}}


\bibitem{Goncu2011}
{Cagatay Goncu} {and} {Kim Marriott}. 2011.
\newblock \showarticletitle{GraVVITAS: Generic Multi-touch Presentation of
  Accessible Graphics}. In {\em Human-Computer Interaction -- INTERACT 2011},
  {Pedro Campos}, {Nicholas Graham}, {Joaquim Jorge}, {Nuno Nunes}, {Philippe
  Palanque}, {and} {Marco Winckler} (Eds.). Springer Berlin Heidelberg, Berlin,
  Heidelberg, 30--48.
\newblock
\showISBNx{978-3-642-23774-4}


\bibitem{Gotzelmann2016}
{Timo G\"{o}tzelmann}. 2016.
\newblock \showarticletitle{LucentMaps: 3D Printed Audiovisual Tactile Maps for
  Blind and Visually Impaired People}. In {\em Proc. ACM SIGACCESS Conference
  on Computers \& Accessibility} {\em (ASSETS '16)}. ACM, 81--90.
\newblock
\showISBNx{978-1-4503-4124-0}
\showDOI{%
\url{http://dx.doi.org/10.1145/2982142.2982163}}


\bibitem{Gotzelmann2014}
{Timo G{\"o}tzelmann} {and} {Aleksander Pavkovic}. 2014.
\newblock \showarticletitle{Towards Automatically Generated Tactile Detail Maps
  by 3D Printers for Blind Persons}. In {\em Computers Helping People with
  Special Needs}, {Klaus Miesenberger}, {Deborah Fels}, {Dominique
  Archambault}, {Petr Pe{\v{n}}{\'a}z}, {and} {Wolfgang Zagler} (Eds.).
  Springer International Publishing, 1--7.
\newblock
\showDOI{%
\url{http://dx.doi.org/10.1007/978-3-319-08599-9_1}}


\bibitem{gual2012visual}
{Jaume Gual}, {Marina Puyuelo}, {Joaquim Llover{\'a}s}, {and} {Lola Merino}.
  2012.
\newblock \showarticletitle{Visual Impairment and urban orientation. Pilot
  study with tactile maps produced through 3D Printing}.
\newblock {\em Psyecology\/} {3}, 2 (2012), 239--250.
\newblock


\bibitem{Holloway2018}
{Leona Holloway}, {Matthew Butler}, {and} {Kim Marriott}. 2018.
\newblock \showarticletitle{Accessible maps for the blind: Comparing 3D printed
  models with tactile graphics}. In {\em Proceedings of the SIGCHI Conference
  on Human Factors in Computing Systems} {\em (CHI '18)}. ACM.
\newblock
\showDOI{%
\url{http://dx.doi.org/10.1145/3173574.3173772}}


\bibitem{Hu2015}
{Michele Hu}. 2015.
\newblock \showarticletitle{Exploring New Paradigms for Accessible 3D Printed
  Graphs}. In {\em Proceedings of the 17th International ACM SIGACCESS
  Conference on Computers \&\#38; Accessibility} {\em (ASSETS '15)}. ACM, New
  York, NY, USA, 365--366.
\newblock
\showISBNx{978-1-4503-3400-6}
\showDOI{%
\url{http://dx.doi.org/10.1145/2700648.2811330}}


\bibitem{kahn2008}
{Peter~H. Kahn}, {Nathan~G. Freier}, {Takayuki Kanda}, {Hiroshi Ishiguro},
  {Jolina~H. Ruckert}, {Rachel~L. Severson}, {and} {Shaun~K. Kane}. 2008.
\newblock \showarticletitle{Design Patterns for Sociality in Human-robot
  Interaction}. In {\em Proceedings of the 3rd ACM/IEEE International
  Conference on Human Robot Interaction} {\em (HRI '08)}. ACM, New York, NY,
  USA, 97--104.
\newblock
\showISBNx{978-1-60558-017-3}
\showDOI{%
\url{http://dx.doi.org/10.1145/1349822.1349836}}


\bibitem{Kane2014}
{Shaun~K. Kane} {and} {Jeffrey~P. Bigham}. 2014.
\newblock \showarticletitle{Tracking@ stemxcomet: teaching programming to blind
  students via 3D printing, crisis management, and twitter}. In {\em
  Proceedings of the 45th ACM technical symposium on Computer science
  education}. ACM, 247--252.
\newblock


\bibitem{kelley1984}
{J.~F. Kelley}. 1984.
\newblock \showarticletitle{An Iterative Design Methodology for User-friendly
  Natural Language Office Information Applications}.
\newblock {\em ACM Trans. Inf. Syst.\/} {2}, 1 (Jan. 1984), 26--41.
\newblock
\showISSN{1046-8188}
\showDOI{%
\url{http://dx.doi.org/10.1145/357417.357420}}


\bibitem{kim2015}
{Jeeeun Kim} {and} {Tom Yeh}. 2015.
\newblock \showarticletitle{Toward {3D}-printed movable tactile pictures for
  children with visual impairments}. In {\em Proceedings of the 33rd Annual ACM
  Conference on Human Factors in Computing Systems}. ACM, 2815--2824.
\newblock


\bibitem{Laranjo2018}
{Liliana Laranjo}, {Adam~G Dunn}, {Huong~Ly Tong}, {Ahmet~Baki Kocaballi},
  {Jessica Chen}, {Rabia Bashir}, {Didi Surian}, {Blanca Gallego}, {Farah
  Magrabi}, {Annie Y~S Lau}, {and} {Enrico Coiera}. 2018.
\newblock \showarticletitle{{Conversational agents in healthcare: a systematic
  review}}.
\newblock {\em Journal of the American Medical Informatics Association\/} {25},
  9 (07 2018), 1248--1258.
\newblock
\showISSN{1527-974X}
\showDOI{%
\url{http://dx.doi.org/10.1093/jamia/ocy072}}


\bibitem{Mayer2003}
{Richard Mayer}, {Kristina Sobko}, {and} {Patricia D.~Mautone}. 2003.
\newblock \showarticletitle{Social Cues in Multimedia Learning: Role of
  Speaker's Voice}.
\newblock {\em Journal of Educational Psychology\/}  {95} (06 2003), 419--425.
\newblock
\showDOI{%
\url{http://dx.doi.org/10.1037/0022-0663.95.2.419}}


\bibitem{McDonald2014}
{Samantha McDonald}, {Joshua Dutterer}, {Ali Abdolrahmani}, {Shaun~K. Kane},
  {and} {Amy Hurst}. 2014.
\newblock \showarticletitle{Tactile Aids for Visually Impaired Graphical Design
  Education}. In {\em Proceedings of the 16th International ACM SIGACCESS
  Conference on Computers \& Accessibility} {\em (ASSETS '14)}. ACM, New York,
  NY, USA, 275--276.
\newblock
\showISBNx{978-1-4503-2720-6}
\showDOI{%
\url{http://dx.doi.org/10.1145/2661334.2661392}}


\bibitem{Moreno2001}
{Roxana Moreno}, {Richard~E. Mayer}, {Hiller~A. Spires}, {and} {James~C.
  Lester}. 2001.
\newblock \showarticletitle{The Case for Social Agency in Computer-Based
  Teaching: Do Students Learn More Deeply When They Interact With Animated
  Pedagogical Agents?}
\newblock {\em Cognition and Instruction\/} {19}, 2 (2001), 177--213.
\newblock
\showDOI{%
\url{http://dx.doi.org/10.1207/S1532690XCI1902\_02}}


\bibitem{Nikitina2018}
{Svetlana Nikitina}, {Sara Callaioli}, {and} {Marcos Baez}. 2018.
\newblock \showarticletitle{Smart Conversational Agents for Reminiscence}. In
  {\em Proceedings of the 1st International Workshop on Software Engineering
  for Cognitive Services} {\em (SE4COG '18)}. ACM, New York, NY, USA, 52--57.
\newblock
\showISBNx{978-1-4503-5740-1}
\showDOI{%
\url{http://dx.doi.org/10.1145/3195555.3195567}}


\bibitem{NFIB2009}
{National~Federation of~the Blind}. 2009.
\newblock The Braille Literacy Crisis in America: Facing the Truth, Reversing
  the Trend, Empowering the Blind.  (2009).
\newblock
\newblock
\shownote{Available from
  \url{https://nfb.org/images/nfb/documents/pdf/braille_literacy_report_web.pdf}.}


\bibitem{Petrie2002}
{Helen Petrie}, {Christoph Schlieder}, {Paul Blenkhorn}, {Gareth Evans},
  {Alasdair King}, {Anne-Marie O'Neill}, {George~T. Ioannidis}, {Blaithin
  Gallagher}, {David Crombie}, {Rolf Mager}, {and} {Maurizio Alafaci}. 2002.
\newblock \showarticletitle{TeDUB: A System for Presenting and Exploring
  Technical Drawings for Blind People}. In {\em Computers Helping People with
  Special Needs}, {Klaus Miesenberger}, {Joachim Klaus}, {and} {Wolfgang
  Zagler} (Eds.). Springer Berlin Heidelberg, Berlin, Heidelberg, 537--539.
\newblock
\showISBNx{978-3-540-45491-5}


\bibitem{Poppinga2011}
{Benjamin Poppinga}, {Charlotte Magnusson}, {Martin Pielot}, {and} {Kirsten
  Rassmus-Gr\"{o}hn}. 2011.
\newblock \showarticletitle{TouchOver Map: Audio-tactile Exploration of
  Interactive Maps}. In {\em Proceedings of the 13th International Conference
  on Human Computer Interaction with Mobile Devices and Services} {\em
  (MobileHCI '11)}. ACM, New York, NY, USA, 545--550.
\newblock
\showISBNx{978-1-4503-0541-9}
\showDOI{%
\url{http://dx.doi.org/10.1145/2037373.2037458}}


\bibitem{Pradhan2018}
{Alisha Pradhan}, {Kanika Mehta}, {and} {Leah Findlater}. 2018.
\newblock \showarticletitle{"Accessibility Came by Accident": Use of
  Voice-Controlled Intelligent Personal Assistants by People with
  Disabilities}. In {\em Proceedings of the 2018 CHI Conference on Human
  Factors in Computing Systems} {\em (CHI '18)}. ACM, New York, NY, USA,
  Article 459, 13 pages.
\newblock
\showISBNx{978-1-4503-5620-6}
\showDOI{%
\url{http://dx.doi.org/10.1145/3173574.3174033}}


\bibitem{rader2014}
{Joshua Rader}, {Troy McDaniel}, {Artemio Ramirez}, {Shantanu Bala}, {and}
  {Sethuraman Panchanathan}. 2014.
\newblock \showarticletitle{A Wizard of Oz Study Exploring How
  Agreement/Disagreement Nonverbal Cues Enhance Social Interactions for
  Individuals Who Are Blind}. In {\em HCI International 2014 - Posters'
  Extended Abstracts}, {Constantine Stephanidis} (Ed.). Springer International
  Publishing, Cham, 243--248.
\newblock
\showISBNx{978-3-319-07854-0}


\bibitem{Reeves2004}
{Leah~M. Reeves}, {Jennifer Lai}, {James~A. Larson}, {Sharon Oviatt}, {T.~S.
  Balaji}, {St{\'e}phanie Buisine}, {Penny Collings}, {Phil Cohen}, {Ben
  Kraal}, {Jean-Claude Martin}, {Michael McTear}, {TV Raman}, {Kay~M. Stanney},
  {Hui Su}, {and} {Qian~Ying Wang}. 2004.
\newblock \showarticletitle{Guidelines for Multimodal User Interface Design}.
\newblock {\em Commun. ACM\/} {47}, 1 (Jan. 2004), 57--59.
\newblock
\showISSN{0001-0782}
\showDOI{%
\url{http://dx.doi.org/10.1145/962081.962106}}


\bibitem{Reichinger2016}
{Andreas Reichinger}, {Anton Fuhrmann}, {Stefan Maierhofer}, {and} {Werner
  Purgathofer}. 2016.
\newblock \showarticletitle{Gesture-Based Interactive Audio Guide on Tactile
  Reliefs}. In {\em Proceedings of the 18th International ACM SIGACCESS
  Conference on Computers and Accessibility} {\em (ASSETS '16)}. ACM, New York,
  NY, USA, 91--100.
\newblock
\showISBNx{978-1-4503-4124-0}
\showDOI{%
\url{http://dx.doi.org/10.1145/2982142.2982176}}


\bibitem{Rossetti2018}
{V. Rossetti}, {Francesco Furfari}, {B. Leporini}, {S. Pelagatti}, {A. Quarta},
  {V. Rossetti}, {F. Furfari}, {B. Leporini}, {Susanna Pelagatti}, {and} {A.
  Quarta}. 2018.
\newblock \showarticletitle{Enabling Access to Cultural Heritage for the
  visually impaired: an Interactive 3D model of a Cultural Site}.
\newblock {\em Procedia Computer Science\/}  {130} (2018), 383--391.
\newblock


\bibitem{Rowell2003}
{Jonathan Rowell} {and} {Simon Ungar}. 2003.
\newblock \showarticletitle{The world of touch: an international survey of
  tactile maps. Part 1: production}.
\newblock {\em British Journal of Visual Impairment\/} {21}, 3 (2003), 98--104.
\newblock
\showDOI{%
\url{http://dx.doi.org/10.1177/026461960302100303}}


\bibitem{Schroeder2013}
{Noah~L. Schroeder}, {Olusola~O. Adesope}, {and} {Rachel~Barouch Gilbert}.
  2013.
\newblock \showarticletitle{How Effective are Pedagogical Agents for Learning?
  A Meta-Analytic Review}.
\newblock {\em Journal of Educational Computing Research\/} {49}, 1 (2013),
  1--39.
\newblock
\showDOI{%
\url{http://dx.doi.org/10.2190/EC.49.1.a}}


\bibitem{Shi2019}
{Lei Shi}, {Holly Lawson}, {Zhuohao Zhang}, {and} {Shiri Azenkot}. 2019.
\newblock \showarticletitle{Designing Interactive 3D Printed Models with
  Teachers of the Visually Impaired}. In {\em Proceedings of the 2019 CHI
  Conference on Human Factors in Computing Systems} {\em (CHI '19)}. ACM, New
  York, NY, USA, Article 197, 14 pages.
\newblock
\showISBNx{978-1-4503-5970-2}
\showDOI{%
\url{http://dx.doi.org/10.1145/3290605.3300427}}


\bibitem{Shi2016}
{Lei Shi}, {Idan Zelzer}, {Catherine Feng}, {and} {Shiri Azenkot}. 2016.
\newblock \showarticletitle{Tickers and Talker: An accessible labeling toolkit
  for 3D printed models}. In {\em Proceedings of the 34rd Annual ACM Conference
  on Human Factors in Computing Systems (CHI'16)}.
\newblock
\showDOI{%
\url{http://dx.doi.org/10.1145/2858036.2858507}}


\bibitem{Shi2017b}
{Lei Shi}, {Yuhang Zhao}, {and} {Shiri Azenkot}. 2017a.
\newblock \showarticletitle{Designing Interactions for 3D Printed Models with
  Blind People}. In {\em Proceedings of the 19th International ACM SIGACCESS
  Conference on Computers and Accessibility} {\em (ASSETS '17)}. ACM, New York,
  NY, USA, 200--209.
\newblock
\showISBNx{978-1-4503-4926-0}
\showDOI{%
\url{http://dx.doi.org/10.1145/3132525.3132549}}


\bibitem{Shi2017a}
{Lei Shi}, {Yuhang Zhao}, {and} {Shiri Azenkot}. 2017b.
\newblock \showarticletitle{Markit and Talkit: A Low-Barrier Toolkit to Augment
  3D Printed Models with Audio Annotations}. In {\em Proceedings of the 30th
  Annual ACM Symposium on User Interface Software and Technology} {\em (UIST
  '17)}. ACM, New York, NY, USA, 493--506.
\newblock
\showISBNx{978-1-4503-4981-9}
\showDOI{%
\url{http://dx.doi.org/10.1145/3126594.3126650}}


\bibitem{Stangl2015}
{Abigale Stangl}, {Chia-Lo Hsu}, {and} {Tom Yeh}. 2015.
\newblock \showarticletitle{Transcribing Across the Senses: Community Efforts
  to Create 3D Printable Accessible Tactile Pictures for Young Children with
  Visual Impairments}. In {\em Proceedings of the 17th International ACM
  SIGACCESS Conference on Computers \&\#38; Accessibility} {\em (ASSETS '15)}.
  ACM, New York, NY, USA, 127--137.
\newblock
\showISBNx{978-1-4503-3400-6}
\showDOI{%
\url{http://dx.doi.org/10.1145/2700648.2809854}}


\bibitem{Taylor2015}
{Brandon~T. Taylor}, {Anind~K. Dey}, {Dan~P. Siewiorek}, {and} {Asim
  Smailagic}. 2015.
\newblock \showarticletitle{TactileMaps.net: A web interface for generating
  customized 3D-printable tactile maps}. In {\em Proc. ACM SIGACCESS Conference
  on Computers \& Accessibility}. ACM, 427--428.
\newblock
\showDOI{%
\url{http://dx.doi.org/10.1145/2700648.2811336}}


\bibitem{vtyurina2018}
{Alexandra Vtyurina} {and} {Adam Fourney}. 2018.
\newblock \showarticletitle{Exploring the Role of Conversational Cues in Guided
  Task Support with Virtual Assistants}. In {\em Proceedings of the 2018 CHI
  Conference on Human Factors in Computing Systems} {\em (CHI '18)}. ACM, New
  York, NY, USA, Article 208, 7 pages.
\newblock
\showISBNx{978-1-4503-5620-6}
\showDOI{%
\url{http://dx.doi.org/10.1145/3173574.3173782}}


\bibitem{Watson2012}
{Alice Watson}, {Timothy Bickmore}, {Abby Cange}, {Ambar Kulshreshtha}, {and}
  {Joseph Kvedar}. 2012.
\newblock \showarticletitle{An Internet-Based Virtual Coach to Promote Physical
  Activity Adherence in Overweight Adults: Randomized Controlled Trial}.
\newblock {\em J Med Internet Res\/} {14}, 1 (26 Jan 2012), e1.
\newblock
\showISSN{1438-8871}
\showDOI{%
\url{http://dx.doi.org/10.2196/jmir.1629}}


\bibitem{wedler2012applied}
{Henry~B Wedler}, {Sarah~R Cohen}, {Rebecca~L Davis}, {Jason~G Harrison},
  {Matthew~R Siebert}, {Dan Willenbring}, {Christian~S Hamann}, {Jared~T Shaw},
  {and} {Dean~J Tantillo}. 2012.
\newblock \showarticletitle{Applied computational chemistry for the blind and
  visually impaired}.
\newblock {\em Journal of Chemical Education\/} {89}, 11 (2012), 1400--1404.
\newblock


\bibitem{Wobbrock2009}
{Jacob~O. Wobbrock}, {Meredith~Ringel Morris}, {and} {Andrew~D. Wilson}. 2009.
\newblock \showarticletitle{User-defined Gestures for Surface Computing}. In
  {\em Proceedings of the SIGCHI Conference on Human Factors in Computing
  Systems} {\em (CHI '09)}. ACM, New York, NY, USA, 1083--1092.
\newblock
\showISBNx{978-1-60558-246-7}
\showDOI{%
\url{http://dx.doi.org/10.1145/1518701.1518866}}


\end{thebibliography}

\end{document}